\documentclass[a4paper,aps,pre,showpacs,twocolumn,superscriptaddress]{revtex4}
\usepackage{graphicx}
\usepackage{rotate}
\usepackage{multirow}
\begin{document}

\title{Impact of community structure on information transfer}

\author{Leon Danon}

\affiliation {Departament de F\'{i}sica Fonamental,Universitat de
Barcelona, Marti i Franques 1, 08028 Barcelona, Spain}
\affiliation{Departament d'Enginyeria Inform\`{a}tica i
Matem\`{a}tiques, Campus Sescelades, Universitat Rovira i Virgili,
43007 Tarragona, Spain}

\author{Alex Arenas}

\affiliation{Departament d'Enginyeria Inform\`{a}tica i
Matem\`{a}tiques, Campus Sescelades, Universitat Rovira i Virgili,
43007 Tarragona, Spain}

\author{Albert D\'{i}az-Guilera}

\affiliation {Departament de F\'{i}sica Fonamental,Universitat de
Barcelona, Marti i Franques 1, 08028 Barcelona, Spain}


\begin{abstract}
The observation that real complex networks have internal structure
has important implication for dynamic processes occurring on such
topologies. Here we investigate the impact of community structure on
a model of information transfer able to deal with both search and
congestion simultaneously. We show that networks with fuzzy
community structure are more efficient in terms of packet delivery
that those with pronounced community structure. We also propose an
alternative packet routing algorithm which takes advantage of the
knowledge of communities to improve information transfer and show
that in the context of the model an intermediate level of community
structure is optimal. Finally, we show that in a hierarchical
network setting, providing knowledge of communities at the level of
highest modularity will improve network capacity by the largest
amount.

\end{abstract} \maketitle

\section{Introduction}\label{sec_intro}

The continuing intensity that accompanies the study of complex networks
has led to  many important contributions in a variety of scientific disciplines
(for a recent review see \cite{Boccaletti06}). Specifically, the study of transport properties of networks is becoming
increasingly important due to the constantly growing amount of
information and commodities being transferred through them. A
particular focus of these studies is how to make the capacity of the
network maximal while minimising the delivery time. Both network
packet routing strategies and network topology play essential parts
in traffic flow in networks.

Traditionally routing strategies have been based on the idea of
maintaining routing tables of the best approximation of the shortest
paths between nodes. In realistic settings, however, the knowledge
that any one of the nodes has about the topology of the network will
be incomplete. So, much of the focus in recent studies has been on
{\em searchability}. In particular, distributed search using only local 
information has been shown to be efficient in spatially embedded 
networks \cite{Kleinberg00,Thadakamalla07}.  Networks with scale-free 
degree distributions  are particularly navigable using local search 
strategies due to the presence of highly connected hubs \cite{Adamic01}.

However, when the number of search problems the network is trying to solve 
increases, it raises the problem of congestion at central nodes. It has 
been observed, both in real world networks \cite{jacobson88} and in 
model communication networks 
\cite{ohira98,tretyakov98,arenas01,sole01,guimera02,Tadic04,Zhao05}, that the
networks collapse when the load is above a certain threshold and the
observed transition can be related to the appearance of the $1/f$
spectrum of the fluctuations in Internet flow data
\cite{takayasu96,Duch06}. 

These two problems, search and congestion, that have so far been
analysed separately in the literature can be incorporated in the
same communication model. Previous work has contributed a collection
of models that capture the essential features of communication
processes and are able to handle these two important issues
simultaneously \cite{arenas01,guimera01,guimera02,guimera02b}. In
these models, agents are nodes of a network and can interchange
information packets along links in the network. Each agent has a
certain capability that decreases as the number of packets to
deliver increases. The transition from a free phase to a congested
phase has been studied for different network architectures in
\cite{arenas01,guimera02}, whereas in \cite{guimera01} the cost of
maintaining communication channels was considered.

The topology of the network also plays a central part in communication
processes. In \cite{guimera02b} the problem of
finding optimal network topologies for both search and congestion
for a fixed number of nodes and links was tackled. It was found that
in the free regime, highly centralised topologies facilitating
search are optimal, whereas in the congested regime decentralised
topologies which distribute the packet load between nodes are
favoured. It has been shown that shortest path routing algorithms
are not optimal for scale free networks due to the presence of
communication bottlenecks \cite{Sreenivasaan06} and several
alternative routing strategies have been proposed to take advantage
of the scale-free nature of complex networks
\cite{Tadic04,Tadic06,Danila06,Danila06b,Toroczkai04}.

On the other hand, many networks found in nature have been observed
to have a modular or {\it community} structure. Communities are
those subsets of nodes that are more densely linked internally than
to the rest of the network. Identifying communities in networks has
become a problem which has been tackled by many researchers in
recent years (see for example \cite{GN,Reichardt04,Guimera05}, and
for reviews see \cite{Newman04,Danon05}). Furthermore, communities
are often organised in a hierarchical way
\cite{guimera02b,Arenas06,Arenas06b,Fortunato07,Sales07}. That is,
large communities are often comprised of several smaller
communities. Despite all these efforts, the impact that community
structure has on information transfer has not been considered.

The aim of this paper is two-fold: firstly, we will investigate the
effect that community structure has on the model of search and
congestion, and secondly we will propose an alternative routing
strategy and demonstrate its impact in the presence of community
structure. In the next section we will describe the model and recall
the most important analytical results. In Section II we will
consider the effect that a modular structure of varying strength has
on the behaviour of the model. We will then show how knowledge of
this community structure can be taken advantage of to improve
transport processes in networks. And in the final section, we give
some concluding remarks.

\section{Communication model}
\label{sec_model}

The communication model considers that the information flowing
through the networks is formed by discrete packets sent from an
origin node to a destination node. Each node is an independent agent
that can store as many packets as necessary. However, to have a
realistic picture of communication we must assume that the nodes
have a finite capacity to process and deliver packets. That is, a
node will take longer to deliver two packets than just one. A
particularly simple example of this would be to assume that nodes
are able to deliver one (or any constant number) information packet
per time step independent of their load, as in the model of decentralised 
information processing in firms of Radner \cite{radner93} and in simple 
models of computer queues \cite{ohira98,tretyakov98,guimera01,sole01}, 
but note that many alternative situations are possible.

In the present model, each node has a certain ability to deliver
packets which is limited. This limitation in the ability of agents
to deliver information can result in congestion of the network. When
the amount of information is too large, agents are not able to
handle all the packets and some of them remain undelivered for
extremely long periods of time. The maximum amount of information
that a network can manage before collapse gives a measure of the
quality of its organisational structure. In this study, the interest
is focused on when congestion occurs depending on the topology of
the network \cite{guimera01}.

The dynamics of the model is as follows. At each time step $t$, an
information packet is created at every node with probability $\rho$.
Therefore $\rho$ is the control parameter: small values of $\rho$
correspond to low density of packets and high values of $\rho$
correspond to high density of packets. When a new packet is created,
a destination node, different from the origin, is chosen randomly in
the network. Thus, during the following time steps $t+1,\,t+2,\ldots
,\,t+T$, the packet travels toward its destination. Once the packet
reaches the destination node, it is delivered and disappears from
the network.

The time that a packet remains in the network is related not only to
the distance between the source and the target nodes, but also to
the amount of packets in its path. Nodes with high loads
---i.e. high quantities of accumulated packets--- will take longer to
deliver packets or, in other words, it will take more time steps for
packets to cross regions of the network that are highly congested.
In particular, at each time step, all the packets move from their
current position, $i$, to the next node in their path, $j$, with a
probability $p_{ij}$. This probability $p_{ij}$ is called the {\it
quality of the channel} between $i$ and $j$. In this paper, we take
the special case that each node is able to send one packet at each
time step. It is important to note, however, that the model is not
deterministic. Here, a packet which is waiting at a particular node,
will be sent with equal probability as any other packet waiting at
the same node.

The packets in the present model have a limited radius of knowledge,
that is, they are able to determine whether a node within a
certain distance $r$ is the destination node. In this case, the packet 
takes the shortest possible route to the destination, otherwise, it travels 
down a link chosen at random. In this paper we set $r=1$, so that only 
nearest neighrbours are recognised. It has been shown in previous work 
that in the free phase, there is no accumulation at any node in the network
and the number of packets that arrive at node $j$ is, on average, $\rho B_j/(S-1)$,
where $B_j$ is the {\em effective betweenness} of node $j$ which is
defined as the fractional number of paths that packets take though node 
$j$ and $S$ is the number of nodes in the network. 
A particular node will collapse when $\rho B_j/(S-1)>1$ and the critical congestion point of the network will be
\begin{equation}
\rho_c=\frac{S-1}{B^*}\, \label{pc}
\end{equation}
where $B^*$ is the maximum effective betweenness in the network, that 
corresponds to the most central node

If the routing algorithm is Markovian, which is the case here, it is possible to 
estimate $B_j$ analytically. The search and congestion process can be 
formulated as a Markov chain, which is dependant on the packet transition
probability matrix. This matrix is derived from the adjacency matrix of the network,
the radius of knowledge $r$, and the search algorithm. Using this formulation, 
$B_j$ of each node can be calculated analytically for any $r$ \cite{guimera02}. In
these cases, the paths the packets take will not be shortest paths. As the radius of
knowledge increases, $B_j$ converges to shortest path betweenness and will be
equal to it when $r$ is greater or equal to the diameter of the network. 

.

\section{Packet dynamics of communication model on networks with community structure}
\sectionmark{Packet dynamics}

The model from \cite{guimera02} can be further exploited to look at
the effects that community structure has on dynamics. To this end we
need to be able to construct networks with controllable community
structure. We choose to use a family of pseudo-random networks since
all other properties (such as node degree and clustering) will be
equivalent to fully random networks. The only thing that we will
vary is the strength of community structure.

First we employ the networks proposed in \cite{NG}. These networks
are comprised of 128 nodes which are split into four communities of
32 nodes each. Pairs of nodes belonging to the same community are
linked with probability $p_{in}$, whereas pairs belonging to
different communities are joined with probability $p_{out}$. The
value of $p_{in}$ is chosen so that the average number of links a
node has to members of any other community, $Z_{in}$, can be
controlled. While $p_{in}$ (and therefore $Z_{in}$) is varied
freely, the value of $p_{out}$ is chosen to keep the total average
node degree, $k$, constant, and set to 16. As $Z_{in}$ is increased
from zero, the communities become better defined and easier to
identify.

To address the question of hierarchical structure we use a
generalisation of the model of generation of networks with community
structure that includes two hierarchical levels of communities as
introduced in \cite{Arenas06}. The graphs are generated as follows:
in a set of 256 nodes, 16 compartments are prescribed that will
represent our first community organisational level. Each of these
sub-communities contains 16 nodes each. Furthermore, four second
level communities are prescribed, each containing four
sub-communities, that is 64 nodes each. The internal degree of nodes
at the first level $Z_{in_1}$ and the internal degree of nodes at
the second level $Z_{in_2}$ are constrained to keep an average
degree $Z_{in_1}+Z_{in_2}+Z_{out}=18$. From now on, networks with
two hierarchical levels are indicated as $Z_{in_1}$ - $Z_{in_2}$,
e.g. a network with 13-4 means 13 links with the nodes of its first
hierarchical level community (more internal), 4 links with the rest
of communities that form the second hierarchical level (more
external) and 1 link with any community of the rest of the network.

As a simple measure of structural efficiency of the network in terms
of packet transport, we can consider the number of packets present
in the network. We allow the dynamics to reach a steady state, which 
we detect by considering the rate at which the number of packets increases 
in the system. Once this rate becomes small, fluctuating around $0$, 
we have reached the end of the transient. It is important to note that when $\rho> 
\rho_c$ the system never reaches a steady state, the mean number of packets 
keeps growing linearly with time, and the rate never becomes very small. We also 
average over several realisations, since the number of packets in the system is subject to statistical fluctuations.

\subsection{Original communication model}

First of all we simulate the dynamics of the model described above, 
in which the packets have no knowledge of the topology of the network 
at the level of community structure. Introducing community structure in the network topology over which the dynamics occur increases the traffic load on the nodes which connect communities. This is in agreement with the finding that cutting links with the highest betweenness separates communities \cite{GN}. It follows that the {\it effective betweenness} of the nodes at each end of the bridge links will also
be increased. As a result, the capacity of the network to deliver
packets is reduced in function of how fuzzy the community structure
is.

\begin{figure}[ht]
\centerline{\includegraphics*[width=0.95\columnwidth]{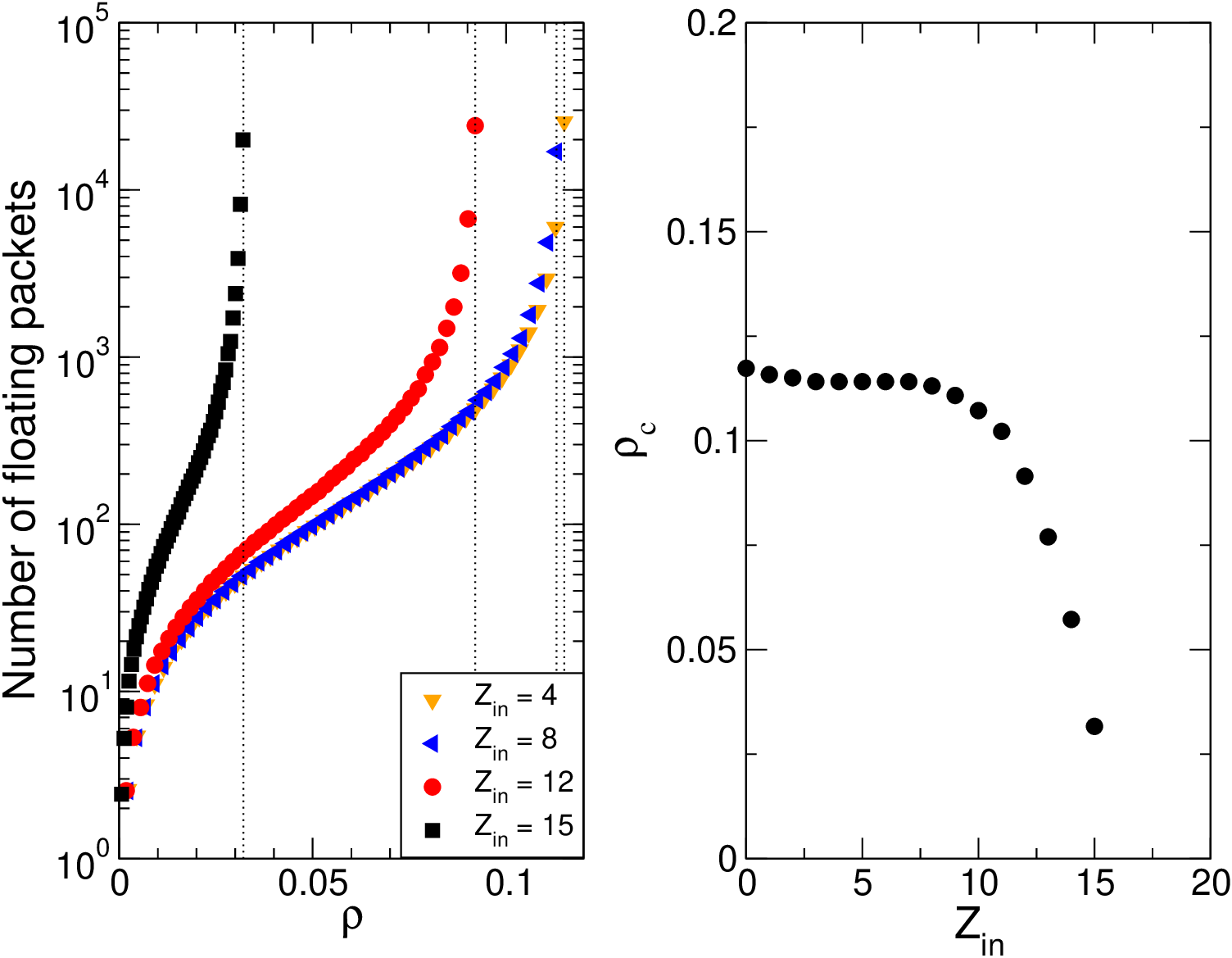}}
\centerline{(a)\hspace{0.38\columnwidth}(b)}
 \caption{
 (colour online) Each point represents the number of packets averaged over 100
 realisations in the steady state of the dynamics.
 (a) Number of floating packets as a function of $\rho$ using
 the original search algorithm in networks with one level of community
 structure. The different colours denote varying levels of community
 strength as controlled by the parameter $Z_{in}$ and the vertical lines
 correspond to the analytical prediction of the onset of congestion
 (Section \ref{sec_model}, equation \ref{pc}).
 (b) Onset of congestion $\rho_c$ for varying $Z_{in}$.
} \label{original_model_coms}
\end{figure}

From Fig. \ref{original_model_coms} we can see that the analytical
calculation from Section \ref{sec_model}, of the onset of congestion
$\rho_c$ agrees very well with the point at which the number of
floating packets diverges. As the strength of community structure is
increased by raising $Z_{in}$, $\rho_c$ is reduced. This seems
logical, since the origin and destination of packets are chosen at
random. It follows that the probability of creating a packet with
both origin and destination in one community is $1/4$. All other
packets will necessarily have to pass through at least one central, "bridge"
node that connects two communities. This leads to an increase in the number of 
packets that pass through bridge nodes, increasing its effective betweenness. 
As a result of receiving a disproportionate amount of packets, these nodes will 
collapse at lower values of $\rho$, leading to a cascade of collapses throughout the network. This effect becomes more and more pronounced as $Z_{in}$ increases, so, the stronger the community structure, the lower $\rho_c$.

\begin{figure}[ht]
\centerline{
\includegraphics*[width=\columnwidth]{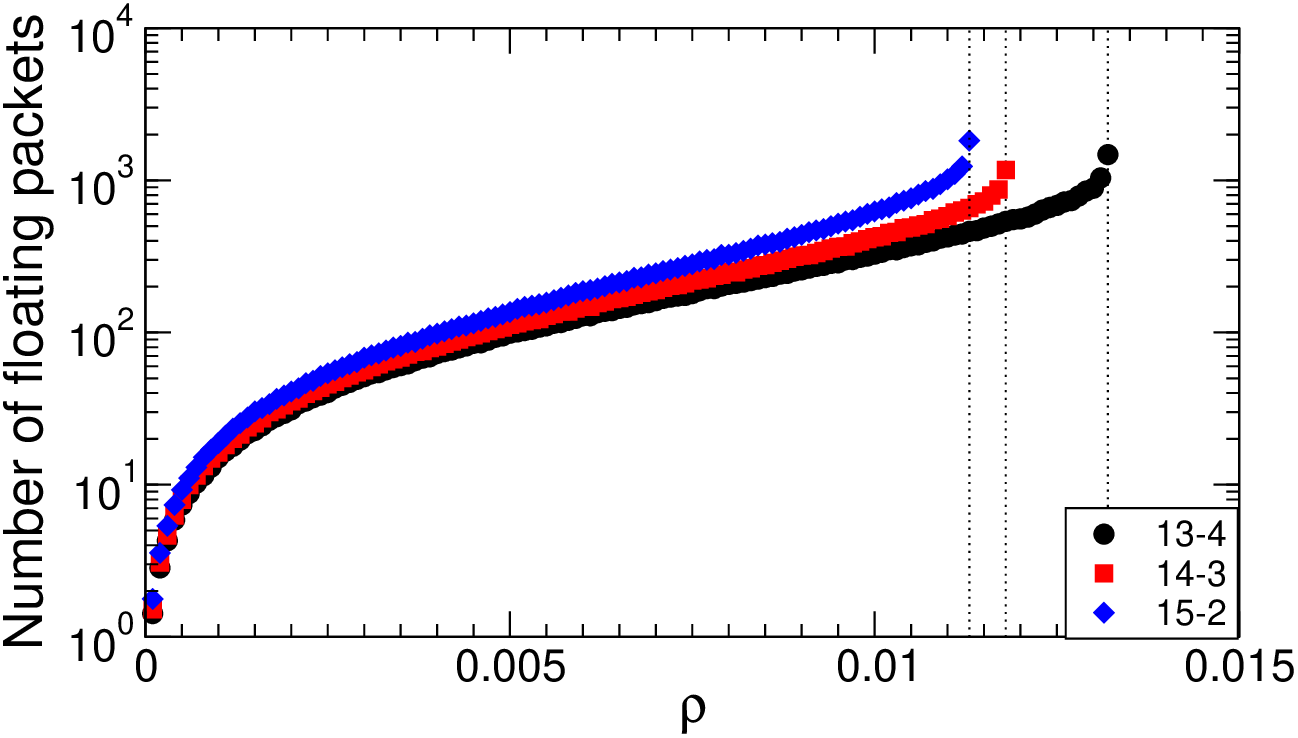}}
 \caption{
     (colour online) Number of floating packets as a function of $\rho$ using
    the original search in networks with hierarchical community structure on two levels.
    The vertical lines correspond to analytical prediction
    of the onset of congestion.
    } \label{original_model_hier}
\end{figure}

In the case of hierarchical networks, we concentrate on three
different network topologies which are particularly instructive,
13-4, 14-3 and 15-2. Once again the analytical calculation
corresponds very well to the point at which the number of floating
packets diverges, see Fig. \ref{original_model_hier}. It is worth
noting that these three networks have almost the same $\rho_c$. This
is due to the fact that the average number of links per node between
communities of size 64 is constant and set to 1. What is varied is
the strength of the intermediate and innermost level of community
structure. In the case of the original communication model, this shows little effect.

\subsection{Modifying the communication model}

Clearly, networks with strong community structure are less efficient
at delivering packets which are oblivious to the underlying
topology. But, what happens when we give the search process some
information about the community structure? To address this question
we propose a simple modification of the way packets are transferred
between nodes.

Let us consider a packet generated at node $i$ in community $c_i$
with destination node $j$ in community $c_j$. At each step in its
path, the packet is given information of the community of
neighbouring nodes. Should the packet destination {\it community} be
the same as that of any of the neighbours of the node that is processing 
the packet, the packet is sent to one of those neighbours, otherwise it 
is sent down a link chosen at random. In this way, packets are able to 
arrive at the destination community without necessarily arriving at the
destination node. The idea is that once within the destination community, 
finding the destination node is easier.

\begin{figure}[!ht]
\centerline{\includegraphics*[width=\columnwidth]{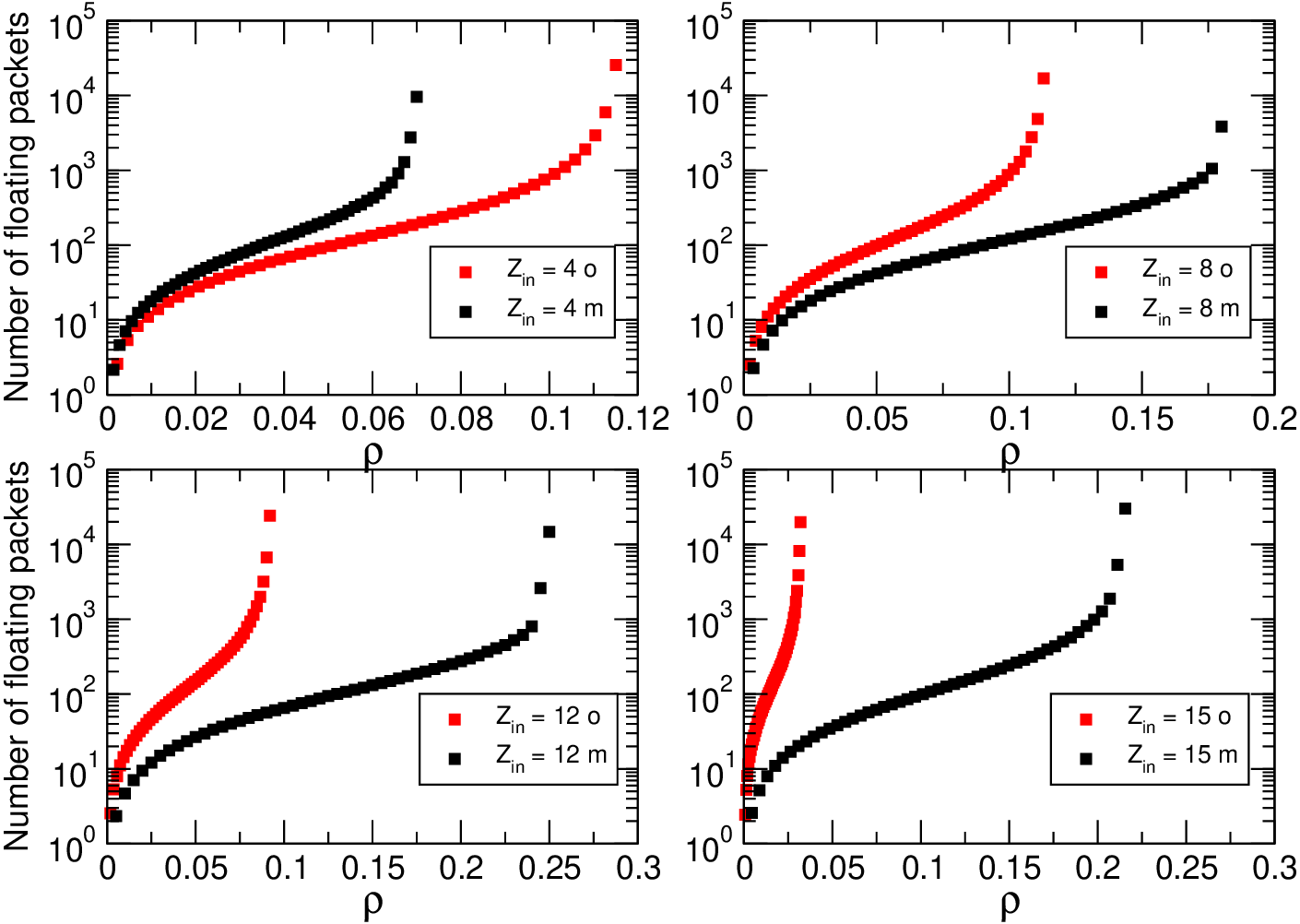}}
\caption{
	 (colour online) Comparison of original search algorithm with the modified
	 for the same networks with community structure. The number of 
	 floating packets is plotted against $\rho$. The four panels depict
	four networks with varying community strength controlled by the
	parameter $Z_{in}$, and the red points show the original search
	algorithm and the black points denote the modified algorithm
	incorporating community information. 
	} \label{fig:comparison_search}
\end{figure}

\begin{figure}[t]
\centerline{
\includegraphics*[width=0.95\columnwidth]{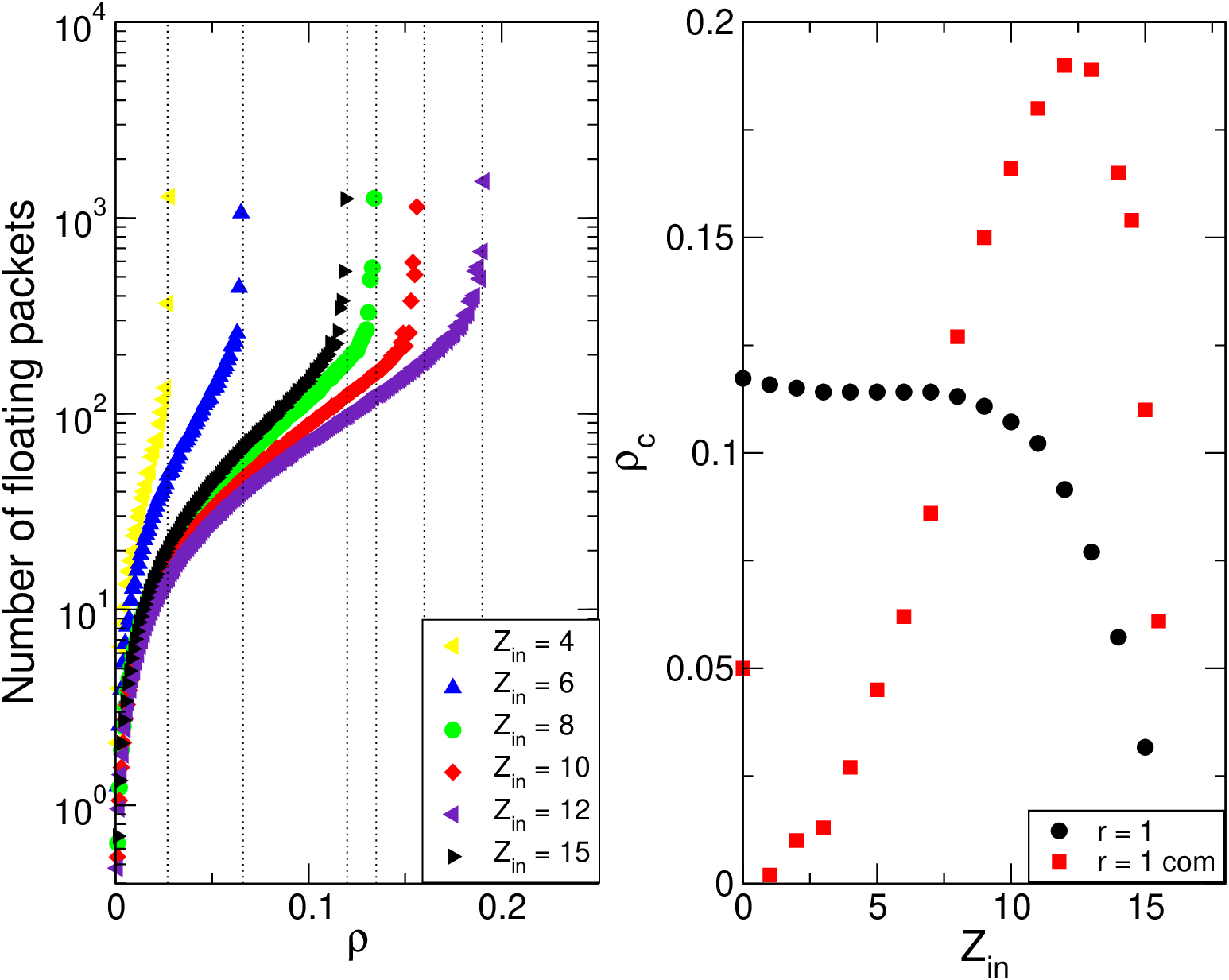}}
\centerline{(a)\hspace{0.38\columnwidth}(b)} 
\caption{ (colour online) Onset of
collapse for the modified search algorithm as applied to networks
with community structure at a single level. (a) Number of floating
packets as a function of $\rho$ for the modified search. Different colours denote 
varying community strengths, controlled by parameter $Z_{in}$ and the vertical 
lines now denote the estimate of the onset of congestion $\rho_c$. (b) $\rho_c$ as 
a function of internal connectivity $Z_{in}$ for both the original model, $r=1$, and 
the modified model, $r=1$ com. 
}
\label{fig:new_algo_search}
\end{figure}

In Figure \ref{fig:comparison_search} we plot the number of floating
packets in the network at the steady state against the packet
injection rate $\rho$. The dynamics are performed on networks with
ad-hoc community structure of varying strength, controlled by the
parameter $Z_{in}$, the average number of links internal to the
community. When $Z_{in}=4$ the network is equivalent to an
Erd\"os-Renyi random graph with 128 nodes and 16 links per node. In
this scenario, the original search algorithm performs much better in
terms of ability to deliver packets. This seems logical: giving
packets information about communities which are not present will not
improve the packet's ability to find the destination node. 
Indeed, for lower values of $Z_{in}$, this information is detrimental to 
efficiency, since the pre-defined partitions of the network actually contain 
{\em fewer} internal links, compared to external ones. In this scenario, 
packets are often sent to regions of the network which are less likely to 
contain the destination node. This is highlighted in Figure \ref{fig:new_algo_search}b. where we see that for very low values of $Z_{in}$ the original search algorithm collapses the network at much higher values of $\rho$ than the modified algorithm.

When the strength of the community structure is increased, the
modified search algorithm improves the efficiency of the network
considerably. For $Z_{in}>8$ \footnote{$Z_{in}=8$ corresponds to
$Z_{out}=8$, at which point most community identification algorithms
are able to detect the correct structure, \cite{Danon05}.}, the
onset of congestion in terms of $\rho$ is considerably higher for
the modified search algorithm, and the same network is much more
efficient at delivering packets for all values of $Z_{in}>8$. In
other words, the modified algorithm is able to find more efficient
routes to deliver packets and the network is able to handle a much
higher load.

In the modified search algorithm, the calculation from
Section \ref{sec_model} (equation \ref{pc}) is still valid, however, the analytic calculation of $B^*$ is more involved than in \cite{guimera02}. Nevertheless 
we can estimate $\rho_c$ of the network by looking at the point where the
number of floating packets diverges. In Figure \ref{fig:new_algo_search}a, 
$\rho_c$ is estimated in this fashion. When the communities are extremely well defined, say $Z_{in}=15$, flow through the network is restricted. So even though the search method of the packets is greatly improved, and they are able to find
the correct community in a short number of steps, flow is restricted
by the formation of bottlenecks at the interface between two
communities. It emerges that an intermediate community structure
strength, $Z_{in}=12$ shows optimal efficiency in terms of $\rho_c$.
This suggests that for the flow to be optimal there must be a
balance between internal strength of communities and connections to
other communities.

For the case of networks with hierarchical community structure as
described above, community information can now be given at two
levels. The packets can be given information about the community
structure on the first level, that is, they are given knowledge
about which community of the four communities of size 64 the
destination node belongs to. From here on, this is denoted as $i=4$.
Alternatively, we can give nodes information on the second level of 
community structure, so that packets know which one of the 16 
communities of size 16 the destination node belongs to, which we 
denote $i=16$.

\begin{figure}[!ht]
\centerline{\includegraphics*[width=\columnwidth]{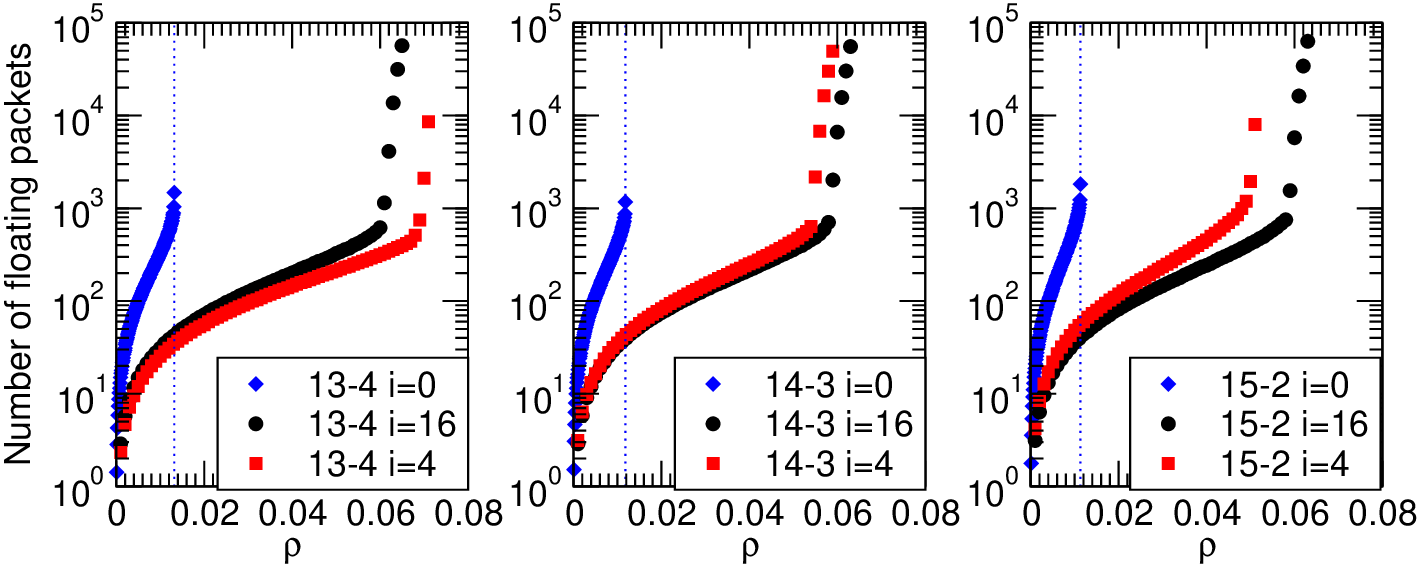}}
\centerline{(a)\hspace{0.3\columnwidth}(b)\hspace{0.3\columnwidth}(c)}
\caption{(colour online) Number of floating packets in networks with hierarchical
community structure. The number of floating packets is show as a
function of $\rho$ for three slightly different hierarchical
networks, (a) 13-4 (b) 14-3 (c) 15-2. For each, the level of
community hierarchy information packets are given is varied between
no information given, $i=0$, information given at the first level
$i=4$ and information given at the second level $i=16$. The vertical
lines represent the analytical calculation as in Section
\ref{sec_model} for the original search algorithm.}
\label{sear:fig_hier_coms}
\end{figure}

Once information about community structure is given to the packets,
the efficiency of the network to deliver these is increased
considerably as in the case of single level community structure. The
level of community information which increases the efficacy of
information flow by the largest amount is dependent on the topology
of the network. Compared with no community information being given
to the packets,  $i=16$ increases the values of $\rho_c$ almost
fivefold, in all three networks. In the case of 13-4, $\rho_c$ is
increased from 0.0132 to 0.064. A stronger community structure at
the second level, 14-3 and 15-2 does not make much of an impact when
$i=16$, with $\rho_c$ being 0.063 for both.

However, when information is given at the intermediate level of community
structure, $i=4$, the differences become more apparent. For the 13-4
configuration, community information at this level favours
information diffusion more, with $\rho_c$ being 0.071, higher than
in the $i=16$ case. However in the case of the 14-3 network, the
opposite is true: giving information at the alternative, $i=16$
level is (marginally) more beneficial. For the 15-2 network, giving
more precise information causes a considerable improvement.

It is interesting to compare these results  with other topological  characterisations 
of complex networks. In particular, the most common measure related to 
community structure is the modularity measure, $Q$, proposed in \cite{NG}
which measures the quality of a particular partition of a particular
network. It is defined as follows:

\[
Q=\sum_i(e_{ii}-a^2_i)
\]
where the element $i,j$ of the matrix $\bf{e}$ represents the
fraction of links between communities $i$ and $j$ and $a_i=\sum_j
e_{ij}$. This value can also be measured at two levels. One is at
the first level of the hierarchy, where nodes are grouped in 4
communities of 64 nodes each, which corresponds to the $i=4$ case.
The other, corresponding to the $i=16$ case, is considering that the
nodes are grouped in 16 communities of 16 nodes each. In the three
networks we are considering, we only vary the strength of the second
level of community structure, so for the $i=4$ case, the value of
modularity remains constant. For the $i=16$ case however, the value
of Q varies with the strength of the second level. For the 13-4
network, the first level of community structure is a better
partition in terms of Q, whereas for 14-3 and 15-2 the second level
is a better partition. See Table \ref{search:hier_table} for values.

\begin{table}[t!]
\centering{\tabcolsep10pt
\begin{tabular}{|c||c|c|c|c|c|}
\hline  
  \multirow{2}{*}{Net} & \multicolumn{3}{c|}{$\rho_c$} & \multicolumn{2}{c|}{Q} \\
  \cline{2-6}
       &$i=0$ & $i=16$ & $i=4$& $i=16$  & $i=4$ \\
\hline
  13-4 & 0.0132 & 0.064 & 0.071 & 0.660 & 0.695\\
  14-3 & 0.0118 & 0.063 & 0.061 & 0.726 & 0.695\\
  15-2 & 0.0113 & 0.063 & 0.053 & 0.771 & 0.695\\
  \hline
  \end{tabular}}
\caption{Table of values of the onset of collapse $\rho_c$ in
hierarchical networks and the values of modularity of the same for
two levels of grouping.} \label{search:hier_table}
\end{table}

For 13-4, where the best partition is found at the first level of
community structure $i=4$, giving packets information about the same
level improves the efficiency of the network more than giving
information at the second level. For 14-3 and 15-2 the opposite is
true: in both cases the best partition is found at the second level,
$i=16$, and the best flow in terms of $\rho_c$ is found when giving
information about the same level. In other words the two coincide.
This means that if communities are organised in a hierarchical
fashion, it is always best to give information at the level where
the maximum modularity is found.

\section{Conclusions}

In this paper we have taken advantage of a model incorporating
search and congestion simultaneously to investigate the impact that
community structure has on information transport. We have shown that
transport is compromised when community structure is introduced in
the network since community structure implies the presence of
bottlenecks. In fact, the better defined the communities are, the
more affected packet transport becomes. We have also shown that
transport can be dramatically improved by providing packets with
information about the community structure. And finally we have shown
that the largest improvements are found when the partition with the
largest modularity is used to provide the information.

This suggests that it is possible to infer {\em a priori} what kind
of information should be given to packets to optimise packet
transport, just by identifying the community structure. By finding
the communities at the level of highest modularity, and providing
information at this level, packet transport appears to be optimal.
The question remains: is this is always the case? It certainly seems
possible to improve information transfer on an arbitrary network
just by providing the search algorithm information about the
community structure at the level of highest modularity. Since
maximising the modularity measure is NP hard \cite{Brandes06}, all
community detection algorithms that depend on maximising modularity
are heuristic approximations and as such different identification
algorithms find different partitions with varying values of optimal
modularity for real networks. The results here suggest that giving
information about the community structure as found by the most
accurate algorithms would be best. But this remains to be shown in
the case of real networks.

\end{document}